# Reply to Saint-Antonin: Low-oxygen-tolerant animals predate oceanic anoxic events

Daniel B. Mills[a,1,2], Lewis M. Ward[a,b,1], CarriAyne Jones[a,c], Brittany Sweeten[a], Michael Forth[a], Alexander H. Treusch[a], and Donald E. Canfield[a]

[a]Department of Biology and Nordic Center for Earth Evolution, University of Southern Denmark, 5230 Odense M, Denmark; [b]Department of Geological and Planetary Sciences, California Institute of Technology, Pasadena, CA 91125; and [c]Department of Microbiology and Immunology, University of British Columbia, Vancouver, BC, Canada V6T 1Z3

It is has been assumed for over half a century that the earliest animals were obligate aerobes with relatively high oxygen requirements. However, the conserved biochemistry and widespread phylogenetic distribution of anaerobic energy metabolism in animals suggests a deep ancestral possession of the genes and enzymes necessary for a facultative anaerobic lifestyle. Additionally, non-bilaterian bodyplans are not expected to require particularly high environmental oxygen levels. This is consistent with experimental evidence demonstrating the low-oxygen tolerance of the sponge *Halichondria panicea*. While it is conceivable that low-oxygen-adapted animals evolved only sometime during the past 541 million years, perhaps in response to oceanic anoxic events, they most reasonably date back to the first animals themselves, as the last common ancestor of animals likely emerged in a relatively low-oxygen world, possessed the genetic means for anaerobiosis, and exhibited a bodyplan conducive to aerobic growth under oxygen levels less than 4% of modern atmospheric saturation.

Saint-Antonin (1) proposes the following scenario: the last common ancestor of living animals (or metazoans) had relatively high oxygen requirements that were not met by the environment until the Neoproterozoic Era (1,000-541 million years ago, or Ma). Subsequently, over the course of the Phanerozoic Eon (541-0 Ma), numerous oceanic anoxic events (OAEs) – episodes of extensive marine anoxia – repeatedly selected for low-oxygen-tolerant marine animals, while eliminating those that retained their high, ancestral oxygen requirements. Therefore, *Halichondria panicea*'s ability to tolerate low oxygen (2) is secondarily derived, and inapplicable to the earliest evolution of animals. We did not consider this alternative interpretation in our original text, and we welcome this opportunity to evaluate it.

In our paper, we argue that *H. panicea*'s ability to endure dissolved oxygen concentrations ≤ 4% of modern atmospheric saturation is likely attributed, at least in part, to its bodyplan. This reasoning is in line with other studies estimating the oxygen requirements of non-bilaterian animals (3, 4). Since all living sponges share the same general organization – or once did in the case of carnivorous sponges – it most likely dates back to the Neoproterozoic divergence of crown-group sponges (5, 6). Indeed, there are a number of unequivocal sponge fossils dating back to the Cambrian Period (7, 8), with some more controversial claims dating back to the Neoproterozoic Era (9-11). Therefore, we can confidently state that the sponge bodyplan predates Phanerozoic OAEs. The fact that our *H. panicea* specimens were collected from well-oxygenated waters, yet endured our experimental conditions, is consistent with a more innate tolerance to low oxygen, potentially based on their conserved morphology.

Although it is unknown whether *H. panicea* underwent any form of anaerobiosis over the course of our experiments, the transcriptomes of several other sponge taxa reveal the presence of genes involved in anaerobic energy metabolism – genes present in other animal lineages and eukaryotes in general (12). Indeed, the facultative anaerobic mitochondria of various metazoan lineages exhibit nearly the same underlying biochemistry and utilize the same core set of genes and enzymes (13). As far as we can tell, there is little to no evidence for lineage-specific acquisitions of these genes in animals (and in eukaryotes in general), which would be expected if they were independently obtained (*e.g.* via horizontal gene transfer) and not inherited through common ancestry. It is most parsimonious, therefore, to posit that the last common ancestor of animals vertically inherited a specific combination of genes for anaerobiosis from its immediate unicellular ancestors, which, in turn, possessed a particular subset of the only 50 or so genes involved in anaerobic energy metabolism in eukaryotes (13).

From likely facultative anaerobic beginnings, various animal lineages, such as our own, have lost their ancestral anaerobic capabilities in the process of adapting to the consistently oxygenated surface conditions of the Phanerozoic biosphere (13). Other animal lineages, primarily those remaining in contact with low oxygen conditions, simply retained them, as low oxygen environments have persisted uninterrupted from the Proterozoic into the Phanerozoic Eon. Therefore, genes associated with anaerobiosis were arguably present in the last common ancestor of metazoans, which lived well before the first OAE, during a time when the oxygen content of the oceans was likely lower and less stable than today's. Indeed, in light of current phylogenetic and biochemical evidence, it is difficult to argue that low-oxygen-adapted and facultative anaerobic animals are specific to the Phanerozoic Eon, especially when the selective pressures for hypoxia tolerance were likely far greater in the Neoproterozoic Era, when animal life actually began.

While a number of animal taxa live in hypoxic and even anoxic environments, many modern animals have a much higher low-oxygen threshold. For example, macrofauna and megafauna sharply decline in both diversity and density in the cores of oxygen minimum zones, demonstrating their general intolerance of low oxygen (14). If OAEs had truly "eradicated marine animals dependent on high oxygen levels," then high-oxygen-demand taxa, such as most marine vertebrates, would have all gone extinct with each OAE. This did not happen for a number of reasons. For one, OAEs likely had oceanographic causes, and were not clearly accompanied by low atmospheric oxygen levels (15). Furthermore, many OAEs were demonstrably regional, as opposed to global in nature, and did not always coincide with extinction events (15). Therefore, OAEs did not necessarily force the ancestors of modern animals to adopt low-oxygen lifestyles. Alternatively, OAEs, where they did occur, may have actually helped preserve the existence of certain animal lineages already equipped to endure low oxygen.

Overall, contrary to Saint-Antonin's proposal, high-oxygen-adapted animals are more likely to be the newcomers – not low-oxygen-tolerant sponges. This is the most parsimonious interpretation if the generally well-oxygenated marine conditions of the Phanerozoic are younger than animal life itself.

[1]D.B.M. and L.M.W. contributed equally to this work.

[2]To whom correspondence may be addressed, E-mail: dmills@biology.sdu.dk